%% file: main.tex
\documentclass[11pt]{article}
\usepackage{PRIMEarxiv}
\usepackage{tabularx}
\usepackage[utf8]{inputenc} %
\usepackage[T1]{fontenc}    %
\usepackage{url}            %
\usepackage{booktabs}       %
\usepackage{amsfonts}       %
\usepackage{nicefrac}       %
\usepackage{microtype}      %
\usepackage{lipsum}
\usepackage{fancyhdr}       %
\usepackage{graphicx}       %
\graphicspath{{media/}}     %
\usepackage{etoolbox}     
\usepackage{makecell} 

\usepackage{soul}

\usepackage{url}
\usepackage[hidelinks]{hyperref}
\usepackage[table]{xcolor}
\usepackage{booktabs}
\usepackage[utf8]{inputenc}
\usepackage[small]{caption}
\usepackage{amsmath}
\usepackage{amssymb}
\usepackage{amsthm}
\usepackage{wasysym}
\usepackage{algorithm}
\usepackage{algorithmic}
\usepackage[switch]{lineno}
\usepackage[table]{xcolor} 
\usepackage{listings}
\usepackage{tikz}
\usepackage{xparse}
\usepackage{enumitem}
\usepackage{soul}
\usepackage{CJKutf8}
\usepackage{tocloft}
\usepackage{float}
\usepackage{subfig}
\usepackage{soul}

\hypersetup{
 colorlinks=true,
 linkcolor=black,
 urlcolor=cyan,
 }
\cftsetindents{subsection}{2em}{3em}

\tikzset{chatstyle/.style={text width=2.8in,rounded corners=2pt}}

\usepackage{setspace}
\usepackage{wrapfig}
\usepackage{adjustbox}
\usepackage{xspace}

\usepackage[most]{tcolorbox}

\usepackage{fontawesome5}
\usepackage{pifont}

\usepackage{multicol}
\usepackage{multirow}
\usepackage{bbding}
\usepackage{circledsteps}
\usepackage{makecell}

\usepackage{fancyhdr}

\usetikzlibrary{mindmap,shadows,calc}
\usepackage[numbers]{natbib}
\tcbset{
    question/.style={
    colback=gray!5!white, 
        colframe=white, 
        boxrule=0.5mm, 
        left=1mm, 
        right=1mm, 
        top=1mm, 
        bottom=1mm, 
        arc=3mm, 
        boxsep=3mm,
        drop shadow={black!50!white}, 
        enhanced,
        overlay={
            \node[fill=cyan!70!black, text=white, rounded corners=1mm, font=\bfseries\scriptsize, inner sep=1mm] 
            at (frame.north west) 
            [xshift=8mm, yshift=-0mm] {Question};
            }
    }
}

\tcbset{
    details/.style={
        colback=gray!5!white, 
        colframe=white, 
        boxrule=0.5mm, 
        left=1mm, 
        right=1mm, 
        top=1mm, 
        bottom=1mm, 
        arc=3mm, 
        boxsep=3mm,
        drop shadow={black!50!white}, 
        enhanced,
        overlay={
            \node[fill=purple!70!white, text=white, rounded corners=1mm, font=\bfseries\scriptsize, inner sep=1mm] 
            at (frame.north west) 
            [xshift=8mm, yshift=-0mm] {Details};
        }
    }
}

\tcbset{
    prompt/.style={
        colback=blue!5!white,         
        colframe=blue!50!black,       
        boxsep=4pt,                   
        arc=5pt,                      
        boxrule=0.8pt,                
        width=\linewidth,
        enhanced,
        drop shadow={blue!30!black},  
        boxed title style={
            colback=blue!20!white,    
            colframe=blue!50!black,   
            boxrule=0.8pt,            
            rounded corners=south,    
            size=small,
            boxsep=3pt,               
            left=3mm,
            right=3mm,
            fontupper=\color{blue!30!black}\bfseries  
        },
        attach boxed title to top center={yshift=-\tcboxedtitleheight/2},  
        top=3mm,
        bottom=3mm,
        coltitle=blue!40!black        
    }
}

\NewDocumentCommand{\heng}
{ mO{} }{\textcolor{red}{\textsuperscript{\textit{Heng}}\textsf{\textbf{\small[#1]}}}}

\newcommand{\squishlist}{
\begin{list}{{{\small{$\bullet$}}}}
{\setlength{\itemsep}{1pt}      \setlength{\parsep}{5pt}
\setlength{\topsep}{-2pt}       \setlength{\partopsep}{0pt}
\setlength{\leftmargin}{2.5em} \setlength{\labelwidth}{1em}
\setlength{\labelsep}{1em} } }

\newcommand{\squishend}{  \end{list}  }
\definecolor{aigold}{RGB}{244,210, 1} 
\definecolor{aigreen}{RGB}{210,244,211} 

\sethlcolor{aigreen}
\definecolor{aired}{RGB}{255,180,181} 
\definecolor{lighterseafoam}{RGB}{194,218,184}

\newtcolorbox{boxJ}{
    sharpish corners, 
    colback = sub, 
    colframe = main, 
    boxrule = 0pt, 
    toprule = 4.5pt, 
    enhanced,
    fuzzy shadow = {0pt}{-2pt}{-0.5pt}{0.5pt}{black!35} 
}

\newtcolorbox{boxK}{
    sharpish corners, 
    boxrule = 0pt,
    toprule = 4.5pt, 
    enhanced,
    fuzzy shadow = {0pt}{-2pt}{-0.5pt}{0.5pt}{black!35} 
}
\definecolor{BerkeleyBlue}{HTML}{003262}
\definecolor{CalGoldHex}{HTML}{FDB515}

\newtcolorbox[auto counter, number within=section]{mysection}[2][]{%
  colback=blue!5!white, colframe=blue!75!black,
  fonttitle=\bfseries, coltitle=black,
  title=My Section~\thetcbcounter: #2,#1}
  
\usepackage{mathptmx}  
\usepackage{datetime}
\usepackage{CJKutf8}
\usepackage{libertinus}

\title{Buy versus Build an LLM: \newline \scalebox{0.8}{A Decision Framework for Governments}}

\begin{document}

\author[1]{Jiahao Lu}
\author[1]{Ziwei Xu}
\author[2]{William Tjhi}
\author[3]{Junnan Li}
\author[4]{Antoine Bosselut}
\author[5,6]{Pang Wei Koh}
\author[1,2]{Mohan Kankanhalli}
\affil[1]{National University of Singapore}
\affil[2]{AI Singapore}
\affil[3]{Salesforce AI Research}
\affil[4]{EPFL}
\affil[5]{University of Washington}
\affil[6]{Allen Institute for AI}

\maketitle
\renewcommand{\thefootnote}{\fnsymbol{footnote}}
\footnotetext[1]{\textit{Corresponding Author: Mohan Kankanhalli (dcsmsk@nus.edu.sg). }}

\begin{abstract}


\begin{spacing}{1.2}
\textbf{Abstract:} Large Language Models (LLMs) represent a new frontier of digital infrastructure that can support a wide range of public-sector applications, from general purpose citizen services to specialized and sensitive state functions.
When expanding AI access, governments face a set of strategic choices over whether to buy existing services, build domestic capabilities, or adopt hybrid approaches across different domains and use cases. 
These are critical decisions especially when leading model providers are often foreign corporations, and LLM outputs are increasingly treated as trusted inputs to public decision-making and public discourse.
In practice, these decisions are not intended to mandate a single approach across all domains; instead, national AI strategies are typically pluralistic, with sovereign, commercial and open-source models coexisting to serve different purposes. Governments may rely on commercial models for non-sensitive or commodity tasks, while pursuing greater control for critical, high-risk or strategically important applications.

This paper provides a strategic framework for making this decision by evaluating these options across dimensions including sovereignty, safety, cost, resource capability, cultural fit, and sustainability.
Importantly, ``building'' does not imply that governments must act alone: domestic capabilities may be developed through public research institutions, universities, state-owned enterprises, joint ventures, or broader national ecosystems.
By detailing the technical requirements and practical challenges of each pathway, this work aims to serve as a reference for policy-makers to  determine whether a buy or build approach best aligns with their specific national needs and societal goals.

The short version of this document is published as an ACM TechBrief at \url{https://dl.acm.org/doi/epdf/10.1145/3797946}, and this document is published as an ACM Technology Policy Council white paper at \url{https://www.acm.org/binaries/content/assets/public-policy/buildvsbuyai.pdf}.

\end{spacing}
\end{abstract}

\newcommand{\coverlogos}[4]{%
\begin{tikzpicture}[remember picture,overlay]
  \node[anchor=south] at ($(current page.south west)!#1!(current page.south east) + (0, #2\paperheight)$)
  {\includegraphics[width=#3]{#4}};
\end{tikzpicture}%
}


\newpage
\newpage
\tableofcontents
\newpage
\begin{CJK*}{UTF8}{gbsn}
\input{sections/introduction}

\input{sections/acquisition_options}
\input{sections/pre_considerations}
\input{sections/framework}
\input{sections/opinionated}
\input{sections/conclusions}

\end{CJK*}
\newpage
\bibliographystyle{define}
\bibliography{reference}
\newpage

\end{document}

%% file: sections/introduction.tex
\section{Introduction}
Large language models (LLMs) are increasingly becoming part of everyday digital interaction, embedded in search, productivity tools, education, and decision support systems. 
At the same time, many governments have begun to view AI, especially LLMs, 
not merely as commercial products, but as strategic technological assets with implications for economic competitiveness, public service delivery, and national resilience.
Because of their general-purpose and scalable nature, LLMs offer governments the potential to support a wide range of public objectives, including improving administrative efficiency, enhancing access to public information, supporting education and workforce training, enabling multilingual service delivery, as well as assisting policy analysis.
As a result, several governments have already taken concrete steps to adopt and deploy national or government-aligned language models, either by buying or building. 
{Importantly, this does not imply a single mandated approach across all domains. In practice, national AI strategies are pluralistic: sovereign models typically coexist with commercial and open-source systems, and are selectively used for specific or public-interest needs.}

Examples of building include Singapore’s government-funded SEA-LION models~\cite{ng2025sealion} which support Southeast Asian languages, and speech-focused LLMs MERaLiON series~\cite{Meralion2025}; Phoenix small 1.0~\cite{pheonix2025} trained to understand materials including government policies and legal documents central to Singapore; Switzerland’s Apertus LLM~\cite{hernandez2025apertus}, which aims to serve the public interest and strengthen Switzerland’s digital sovereignty;  UAE-backed efforts such as TII’s Falcon~\cite{almazrouei2023falcon}, alongside other regionally optimized Arabic models including Jais~\cite{sengupta2023jais} and Command R7B Arabic~\cite{alnumay2025command}, which target improved alignment with local dialects and cultural considerations;
and Malaysia launched their fully homegrown multimodal LLM ILMU AI~\cite{ilmu2025} which outperforms global models in Malay-language benchmarks. 
Additionally, recent initiatives in France and Germany that partner with European firms such as Mistral AI and SAP SE to develop sovereign AI systems for public administration~\cite{JointPress2025Join}.

Compared to building sovereign models, examples of governments purchasing LLM services tend to be less frequently disclosed.
Buying is more commonly associated with targeted, domain-specific applications rather than national, general-purpose deployments.
For instance, the U.S. government has signed contracts with OpenAI for use within the Department of Defense~\cite{defense2025contract}.
Similarly, the Australian government entered into a 12-month agreement with OpenAI for the “provision of AI” services~\cite{treasury2025contract, australia2025contract}. 
Details of such contracts are often undisclosed, given confidentiality provisions and the sensitivity of many public-sector contexts. 
It is therefore reasonable to assume that more procurement arrangements exist than are publicly documented.

Beyond a pure build-or-buy dichotomy, a hybrid pathway has also emerged as a viable option. 
Vietnam exemplifies this approach by incrementally assembling a sovereign AI stack which combines domestic platforms, sovereign cloud infrastructure, and high-performance compute, while partnering with foreign enterprises such as NVIDIA, NTT Data, and Huawei.~\cite{vietnam2025news}
In parallel, {collaboration with the Canadian AI institute Mila supports knowledge transfer and long-term talent development~\cite{vietnam2025news}.}
This model illustrates how governments can pursue adaptive capacity building without committing exclusively to either full internal development or complete reliance on external vendors.

As reflected by the examples above, the ``government'' in the context of this paper refers not only to central ministries or agencies, but also to the state-aligned institutions which performs execution, including: (i) the actual government bodies; (ii) state-backed research and public-sector entities (e.g., AI Singapore or TII of UAE's); and (iii) partnerships with private firms where the government has sufficient contractual, regulatory, or strategic leverage to ensure national objectives take precedence (the relevant firms vary by country).

For governments that have not yet taken such steps of deploying LLMs, a fundamental question naturally arises when considering the provisioning of AI capabilities to public agencies or citizens: should a government build its own LLM, or buy access to existing commercial models? 
This question is non-trivial. 
Leading LLM providers are often foreign companies operating under different legal jurisdictions, governance norms, and commercial incentives, making the choice strategically consequential for many states.

Despite the importance of this decision, we observe a lack of structured, technically informed guidance from the research community that is tailored to public-sector decision-making. 
While there are useful discussions offering buy-or-build guidance for enterprises~\cite{IMDA2025playbook, tryolabs2024navigating}, these frameworks are largely designed for commercial contexts and do not fully address the unique constraints and responsibilities faced by governments.
This paper aims to help fill this gap by proposing a structured, neutral evaluation framework to help governments reason about the build–buy decision across multiple strategic dimensions including sovereignty, cost, capability, risk, and sustainability. 
Rather than advocating a direct answer, we aim to surface the key trade-offs and constraints that shape different choices, and to provide policymakers with a set of analytical lenses and empirical reference points to support informed decision-making in diverse national contexts. 
The guidance presented here is shaped in part by our involvement and experience in public-sector LLM efforts such as SEA-LION~\cite{ng2025sealion} and  Apertus~\cite{hernandez2025apertus}, where the practical realities of buy-versus-build decisions have been debated and tested.

%% file: sections/acquisition_options.tex
\section{Acquisition Pathways for Language Models}
\label{sec:acquisition_options}

In this section, we first examine the various pathways for acquiring and managing language model capabilities. 
This choice is not a simple binary decision. 
Governments can acquire models in different ways, with different levels of control, responsibility, and ownership.

In practice, governments may employ a phased approach: buy first for rapid deployment, while simultaneously investing in talent and infrastructure for a future build. 
In this sense, buy and build can be viewed as actions taken across different time frames along a capacity-building trajectory (see Section~\ref{sec:evolving_landscape} for detailed discussions of temporal and path-dependent considerations in the evolving cost-capability landscape). 
For clarity, our taxonomy below distinguishes acquisition pathways within the same decision horizon.

\subsection*{Buy an LLM}

\begin{enumerate}
    \item \textbf{Purchase through API calling}. 
    The most accessible entry point is connecting to a model via a cloud interface managed by a third party, and the government pays for the number of tokens processed. 
    This approach offers the fastest time to deployment and requires almost no internal hardware. 
    However, it relies heavily on the service provider for security and availability, and data must leave internal government networks for processing.
    This raises concerns regarding information privacy, long-term vendor lock-in, and the potential exposure or reuse of sensitive national data in the training of foreign models.

    \item \textbf{Purchase model instances or private deployments}.
    Rather than paying per request, some vendors allow organizations to buy a dedicated instance or a license for a trained model (e.g. AWS Bedrock~\cite{AWS2025bedrock} or Cohere private deployment~\cite{cohere2025private}).
    This option could be more cost-effective for high-volume use and provide a more stable environment where the model version does not change unexpectedly. 
    It often includes stronger privacy guarantees since the instance can sometimes be hosted within a government approved cloud perimeter. 
    While this requires more effort to configure than a simple API, it eliminates the variable cost spikes associated with public API services and ensures that sensitive data never leaves a controlled environment. 
    However, its cost-effectiveness relative to API-based access depends strongly on usage volume: it can be more economical for sustained high workloads, but more expensive under low usage.
\end{enumerate}

\subsection*{Build an LLM}

\begin{enumerate}
    \item \textbf{Build from scratch}.
    This pathway involves training a model using raw data and massive computational clusters. 
    It provides the highest level of data sovereignty and allows the government to control every aspect of the training process, including the values and knowledge embedded in the system. 
    Nevertheless, building from scratch can potentially be prohibitively resource-intensive and demands significant investments in data, computing infrastructure, energy supply and engineering talents. Of course, resource requirements also grow rapidly with model size, often increasing super-linearly (even exponentially) as parameter counts and training compute scale.
    
    \item \textbf{Tuning based on open-source models}.
    A common alternative is to take a powerful open-source base model and further adapt it on specific data or tasks. 
    This adaptation can include continued pre-training~\cite{gururangan2020don, ke2024continual, roth2024practitioner} or mid-training~\cite{tu2025survey, mo2025mid} on national or domain-specific corpora, full or parameter-efficient fine-tuning (e.g., LoRA~\cite{hu2022lora}) and instruction tuning~\cite{mishra2022cross, wei2022finetuned, ouyang2022training} or safety tuning~\cite{zong2024safety, bianchi2024safety} on curated preference data.
    In this way, a government can inherit the general capabilities of a globally developed model while injecting local knowledge, norms, and constraints. 
    Compared to training from scratch, this approach is far more cost-efficient, yet still allows the government to retain full control over the adapted model weights and deployment. 
    It should be noted, however, that most ``open-source’’ models are in practice open-weights models, so not all details of their original training process or data are fully transparent.
\end{enumerate}

\noindent
Importantly, ``building'' in this context refers to strategic ownership and control over model development and evolution, rather than the specific legal form of the executing entity.
In practice, a government’s decision to build a language model does not imply a single organizational form. 
Such efforts can be undertaken through multiple execution pathways. 
First, a model may be developed directly within a government ministry or agency, with full in-house control over development and deployment. 
Second, governments may build LLMs through publicly funded research institutes, non-profit organizations, or national laboratories that operate with a public-interest mandate while retaining technical autonomy. 
Third, governments may choose to build through corporatized entities, including state-owned enterprises, government-controlled joint ventures, companies established with public funding, or even private-sector firms over which the government can exercise effective governance influence to ensure alignment with national objectives. 
These pathways differ primarily in governance structure and operational flexibility, rather than in the strategic intent to build sovereign capabilities.

\begin{figure}[h!]
  \centering
  \includegraphics[width=\linewidth]{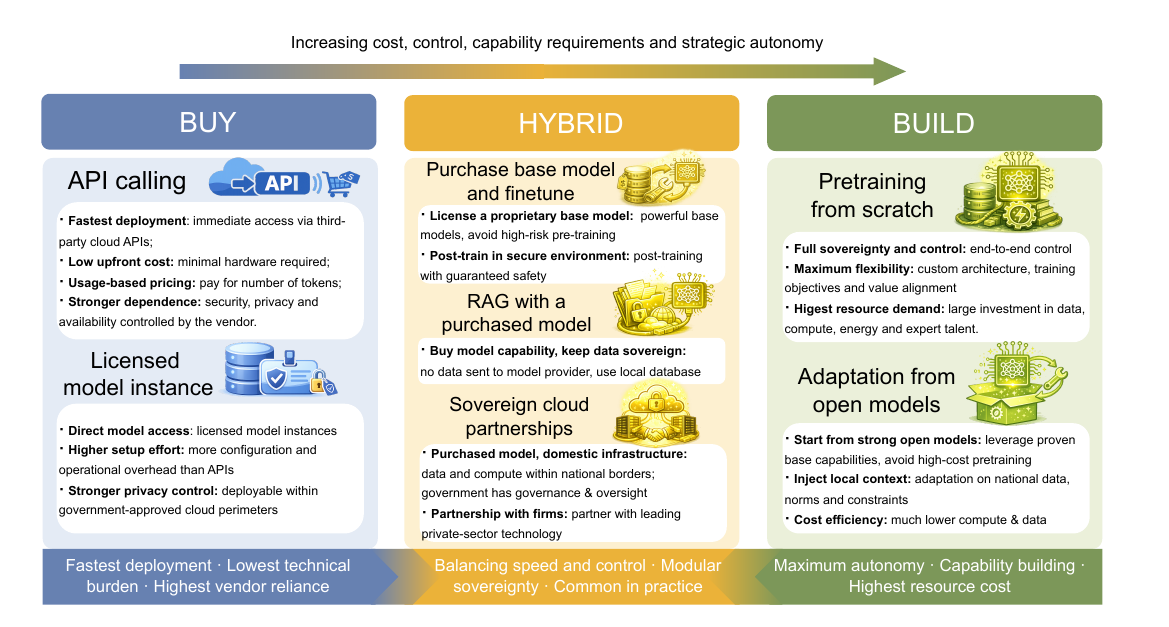}
  \caption{\textbf{Acquisition pathways for language models}, spanning buy, hybrid and build approaches, with progressively higher cost and capability requirements but increasing levels of control and strategic autonomy.}
  \label{fig:acquisition_pathways}
\end{figure}

\subsection*{Hybrid options}
\begin{enumerate}

    \item \textbf{Purchase pre-trained base models and fine-tune}.
    This pathway involves licensing a mature proprietary foundational model from a leading technology firm and performing post-training within a secure government environment.
    By starting with a base model that powers established commercial products, a government avoids the high risk and experimental costs of initial pre-training.
    Once the base model is secured, the government can execute the post-training phase on its own sensitive or local datasets.

    \item \textbf{Retrieval Augmented Generation (RAG) with a purchased model}.
    This is a widely adopted ``sovereign data'' strategy for governments, allowing them to use high capability models (buy) to answer questions based strictly on local, private files (build sovereign database). A specific example is SkillsFuture Singapore (SSG) which uses GraphRAG to analyze $30,000$ customer cases quarterly~\cite{SSG2025solving}. By buying the reasoning power of an LLM but keeping their sensitive data in a private graph database, they process data $62.5$ times faster without leaking citizen information to the model provider.
    Importantly, this is different from connecting public APIs (e.g., ChatGPT or Gemini) directly to sensitive databases - instead, both inference and data access occur within a controlled environment.

    \item \textbf{Joint ventures and sovereign cloud partnerships}. 
    This is a geopolitical hybrid where a country partners with a global tech company to build a ``walled garden'', by buying the LLM service but keeping the hardware and data stay within national borders.
    For instance, in November 2025, the governments of France and Germany announced a strategic partnership with Mistral AI and SAP to deploy ``AI-native sovereign solutions'' for public administration~\cite{SAP2025Ecosystem, JointPress2025Join}.
    It uses Mistral’s high-performance models and SAP's enterprise infrastructure, but the entire system is hosted in European data centers and overseen by a board of European nationals.
    This allows the states to reclaim technological control while still benefiting from cutting-edge private sector research.
    
\end{enumerate}

{Importantly, regardless of the chosen pathway, governments should pursue diversification to improve resilience across the LLM supply chain. This includes diversifying software vendors, hardware dependencies including GPUs, technical experts, and energy suppliers. Such multi-layer diversification reduces concentration, lock-in risks, and single points of failure.}

Figure~\ref{fig:acquisition_pathways} summarizes the buy-hybrid-build acquisition options for governments, and highlights the associated trade-offs in cost, control, capability requirements and strategic autonomy. 
\newpage

%% file: sections/pre_considerations.tex
\section{Pre-Decision Considerations}

Before evaluating the specific trade-offs between building or buying an LLM, decision-makers should first examine a set of foundational considerations that arise in either pathway. 
These considerations manifest differently across countries and they can materially shape or constrain what a viable build or buy strategy looks like in practice. 
Accordingly, rather than advocating a single preferred option, this section highlights key pre-decision factors and show some data points and case studies to help readers assess how these factors apply in their national context and how they can inform subsequent build/buy choices.

\begin{figure}[h]
  \centering
  \includegraphics[width=\linewidth]{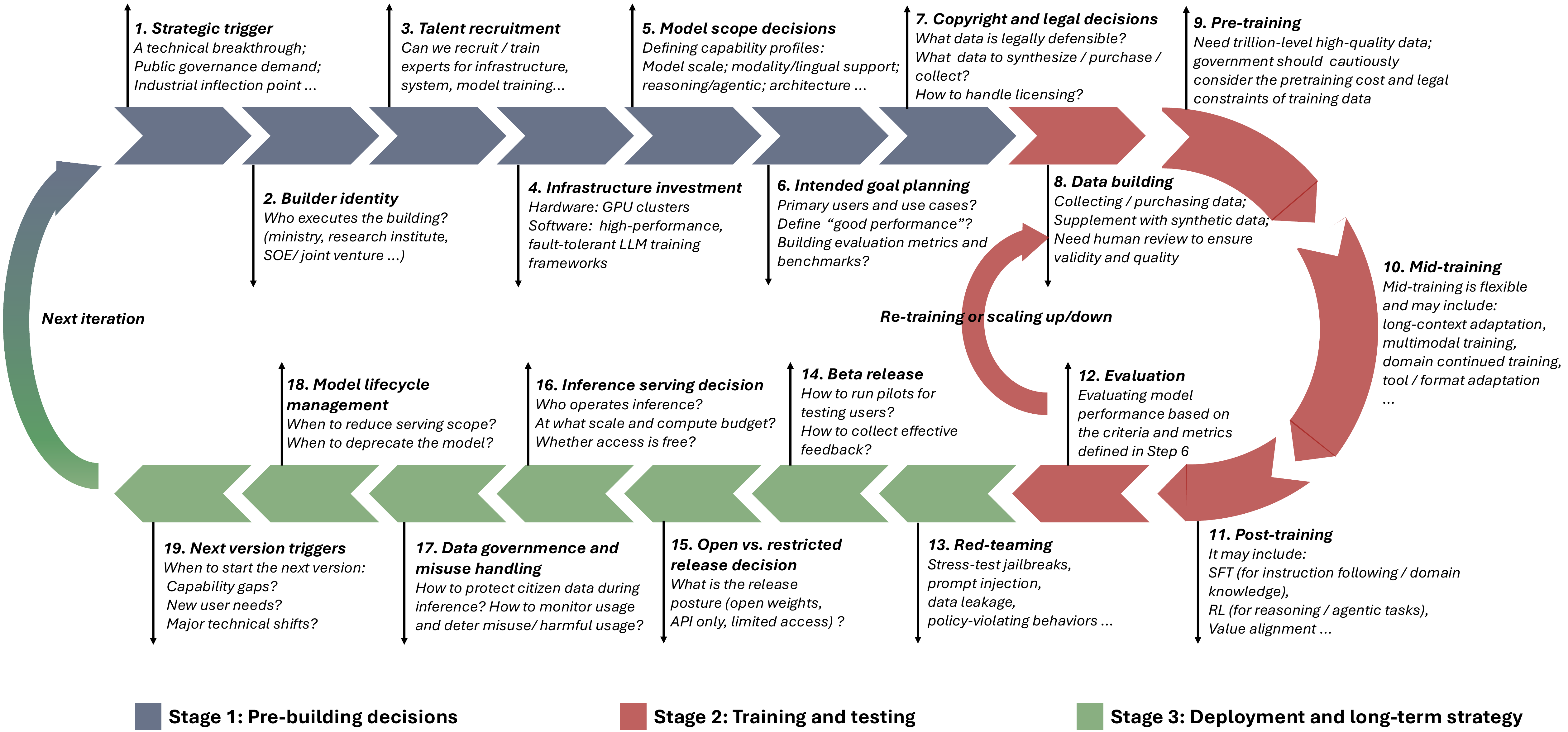}
  \caption{\textbf{Critical decision points across the sovereign LLM building lifecycle}. This figure illustrates key strategic, technical and operational decision calls that governments face when building a sovereign LLM.}
  \label{fig:decision_points}
\end{figure}

\subsection{Categories of Language Models}
\label{sec:language_model_categories}

In practice, large language models (LLMs) come in many variants, and governments should begin with articulating their intended deployment and application scenarios to determine the capability profile that best fits actual needs.

We briefly highlight several key axes along which language models differ, each shaping the suitability of a model for government deployment:

\begin{itemize}
    \item \textbf{Monolingual vs. Multilingual}: 
    For multi-lingual societies, the breadth of language support is a matter of inclusive governance and equitable service delivery.
    While \textit{buying} provides immediate access to globally dominant languages, commercial models might often under-perform in minority languages or dialects due to data scarcity.
    \textit{Building} allows governments to prioritize their specific linguistic landscape, though it needs the costly curation of low-resource data.
    The gap of multilingual support has motivated region-specific efforts such as the SEA-LION family of LLMs~\cite{ng2025sealion}, which has been developed explicitly to support a wide range of Southeast Asian low-resource languages, including Filipino, Burmese, Lao, Khmer, Tamil and others which are poorly covered by many global models. 
    Similarly, Indian initiatives led by AI4Bharat and Sarvam AI have produced Indic-focused models such as Airavata~\cite{gala2024airavata}, a Hindi instruction-tuned LLM and Sarvam-1~\cite{sarvam2024}, a multilingual model tailored to major Indian languages, further illustrating how locally developed LLMs can better serve diverse linguistic landscapes than generic global models.

    \item \textbf{Text-only vs. Multi-modal}: 
    Multi-modal models expand a system’s input space to include image, audio, and video, allowing for more comprehensive data processing. A nation's culture is usually captured in text, images, sounds and videos.
    However, incorporating more modalities adds technical overhead by increasing architectural complexity and resource demands.

    \item \textbf{Reasoning vs. Non-reasoning}: 
    Reasoning and non-reasoning models differ in their capacity for long thinking.
    Compared with non-reasoning models, reasoning models utilize more test-time compute to explore various paths and strategies before producing a final answer~\cite{OpenAI-o1, guo2025deepseek}.
    This advanced capability significantly improves performance on complex tasks that require deep deliberation, such as legal interpretation or strategic planning.
    However, these benefits come with higher computation costs. 
    Governments must weigh the necessity of such depth against the speed of standard models.
    An emerging alternative is the use of adaptive systems that dynamically allocate a thinking budget or toggle reasoning capabilities based on the complexity of the specific request.~\cite{GPT-5.1, Gemini-thinking, xu2025scalable, li2025steering, wu2025arm}.

    \item \textbf{Conversational vs. Agentic}: 
    Conversational models produce text or media responses within a dialogue. 
    Agentic models go further by interacting with external environments to execute tasks. 
    These agents can perform functions such as browsing websites or databases, calling external tools or managing resource allocation~\cite{yao2022react, schick2023toolformer}.
    While agentic systems automate complex workflows and boost productivity, they also introduce significant security risks~\cite{greshake2023not, fang2024llm}. 
    Granting models the power to act can lead to unintended system changes or data breaches. 
    Governments must therefore decide how much autonomy to allow. Agentic workflows can be deployed with strict human-in-the-loop controls, approvals, and audit mechanisms, balancing efficiency gains against residual security risk.

    \item \textbf{General-purpose vs. Domain-Specific}: 
    General purpose models provide broad utility but often lack the precision required for specialized tasks like legal interpretation or regulatory compliance. 
    Domain specific models focus on narrower applications, such as in healthcare or in finance.   
    This specialization avoids the cost of training on irrelevant data and reduces hallucinations caused by conflicting information~\cite{kazlaris2025illusion, gautam975impact}. 
    However, success depends on high quality data and domain expertise, which remain significant hurdles for many government organizations.

    \item \textbf{Open-source vs. Closed-source}: 
    The degree of openness varies across different levels, including model parameters, training code, and source datasets. 
    For an open-source model, governments can leverage global community contributions for testing, security auditing, and continuous upgrades. 
    This transparency often builds public trust. 
    However, the potential risks include malicious fine-tuning~\cite{halawi2024covert} or unauthorized use for illegal purposes~\cite{Anthropic2025Detecting}. Moreover, making pre-training data open-source undermines the unique advantage of that country's models.  
    In contrast, a closed-source model allows the government to maintain total control over its intellectual property and security perimeter. 
    While this closed-source option protects sensitive architectural details, it requires the government to handle all maintenance and improvements internally without the benefit of community innovation.
    
    \item \textbf{Model Size}: 
    Model size is a primary driver of both capital and operational costs. 
    For those considering a build strategy, larger models require significant capital for high performance hardware and specialized talent. 
    Small models are more accessible for internal development and can be optimized on modest infrastructure.
    When evaluating a buy strategy, cost is typically tied to usage volume.
    Larger models often command higher pricing per token, which can lead to high variable costs for frequent tasks.
    Governments must consider whether the extensive capabilities of a massive model justify the recurring fees or the substantial initial investment required for a private build. 
    In contrast, smaller models may provide sufficient performance for specific administrative tasks while keeping both procurement and maintenance costs manageable.
\end{itemize}

These axes jointly define the target of any build-or-buy decision.
Together, they form a multi-dimensional capability profile that a government must map against its specific operational mandates, budget constraints, and risk tolerance.

\subsection{User Adoption and Retention}

When considering whether to build or buy, it is not enough to ask what is technically feasible. Governments should also consider whether people will actually choose to use the deployed language model, and whether they will continue to use it over time.

In a landscape where users have access to many commercial and open-source language models, a government-backed system must compete for attention and trust. Either buying or building an LLM requires substantial effort and resources; if usage remains low, the strategic and economic value of those investment will be limited. 
According to the Menlo Ventures 2025 State of Generative AI report~\cite{Menlo2025Report}, the three market leaders Anthropic, OpenAI, and Google collectively capture 88\% of the enterprise LLM API market share. This high level of concentration indicates that user selection is heavily long tail distributed. 

How to design services, interfaces and incentives that encourage user engagement is beyond the scope of this paper. Nonetheless, expected adoption and retention should play a significant role in the buy–build decision. Before committing to a particular option, decision-makers should consider questions such as: How likely are target users to adopt this model? Will they find it convenient, trustworthy, and useful enough to use regularly, rather than turning to alternative models?

\subsection{Usage Scenarios}

\begin{figure}[h]
  \centering
  \includegraphics[width=0.9\linewidth]{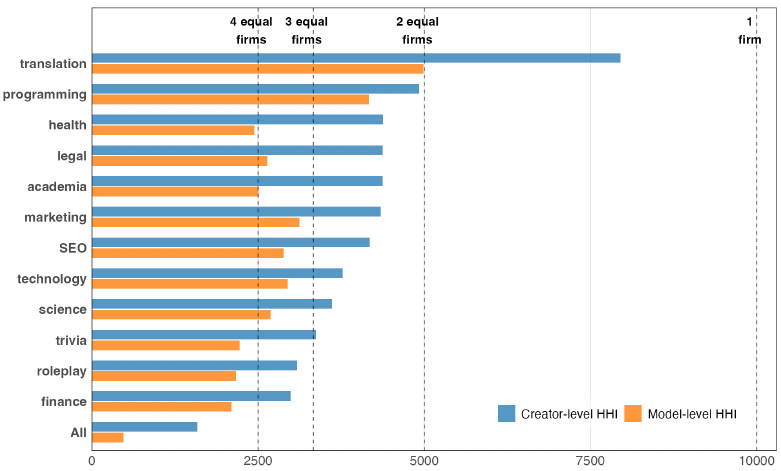}
  \caption{\textbf{Market concentration across application categories measured by the Herfindahl–Hirschman Index (HHI).} HHI is a standard antitrust metric, ranging from 10,000 for a monopoly to 2,500 for a market with four equally sized firms. Categories on the y-axis represent \textbf{use-case categories} reported by OpenRouter~\cite{openrouter}, a marketplace and API gateway that routes developer traffic to selected LLM providers. Market shares are computed from total token usage on OpenRouter from January to December 2025. The figure reports concentration at the \textbf{creator level} (blue; the model developer/company, e.g., OpenAI, Anthropic) and the \textbf{model level} (orange; specific model variants). Adapted from Figure 19 of~\cite{demirer2025emerging}.}
  \label{fig:HHI}
\end{figure}

An institution should also carefully consider the intended usage scenarios of the deployed LLMs. Expected use cases directly shape what capabilities are needed and how it should be developed.

First, usage scenarios influence the \textit{capabilities} required. For example, a system might be designed for \textit{conversational} interaction to answer routine public inquiries, or it may require \textit{agentic} proficiency to autonomously utilize external tools, query databases, and execute complex multi-step workflows.
Second, use cases determine the \textit{scope of training and adaptation data}. For example, deciding whether to include domain-specific corpora, such as tax codes, medical guidelines, or legal precedents.
Third, usage scenarios affect \textit{safety and auditing requirements}. Higher-risk scenarios demand stricter alignment, logging, human-in-the-loop oversight, and external auditing.

Beyond internal requirements, usage scenarios should also be informed by the broader market structure of existing LLM services. 
As illustrated by Figure~\ref{fig:HHI}, LLM  usage across application categories exhibits varying degrees of market concentration, with some domains (e.g., translation, programming, legal and health-related tasks) dominated by a small number of commercial providers. 
Such concentration not only signals potential strategic dependence, but also reveals where user expectations, workflows, and switching costs are already shaped by incumbent models.
For governments, this analysis can help identify where a sovereign model is unlikely to displace entrenched providers, and where instead it may be better positioned as a complementary alternative, by focusing on domains with unmet public-sector needs, lower commercial incentives, or strong requirements for trust, localization, and accountability.
In this sense, usage scenarios are not chosen in isolation, but emerge at the intersection of public objectives and existing market realities.

Therefore, before deciding whether to build or buy, policy-makers should first clarify which scenarios they intend to support and how critical or sensitive those scenarios are. This assessment should be informed by the existing market structure, as governments are unlikely to compete effectively in highly concentrated domains dominated by commercial providers. Instead, sovereign models may deliver greater value by positioning themselves in complementary roles where public trust, localization, accountability, or long-term accessibility are paramount.

\subsection{Life-Cycle Considerations}
The buy--build decision should be evaluated across the full life cycle of the deployed system rather than treated as a one-time procurement or engineering project. Language models evolve rapidly as new architectures, training techniques, safety methods, and capabilities emerge; as a result, a model that is adequate today may become outdated or operationally suboptimal within a few months.

For a build strategy, life-cycle planning centers on the government's ability to continuously retrain, upgrade, and maintain the model as data distributions, policy requirements, and user expectations change. 
Additionally, sustaining developer capability is also important, as training methodologies, tools and best practices change rapidly over time.
For a buy strategy, governments must assess the longevity of vendor offerings, the stability of pricing and service terms, and the feasibility of switching providers or migrating workloads as needs evolve. 
In both cases, long-term planning should explicitly account for model obsolescence, migration pathways, and institutional readiness to revisit the buy--build decision as both technology and governance environments change.

Across major commercial providers (e.g., OpenAI, Anthropic, and Google), model versions are frequently superseded or deprecated within months to roughly a year, which illustrates how quickly the capability baseline and user expectations can shift.
Therefore, even if government-provided LLMs do not need to update at the same pace, decision-makers should still plan for periodic upgrades and migration windows to prevent the deployed system from becoming functionally outdated and gradually losing users to alternatives.

\subsection{Technical and Resource Demands of Building}
\input{figures/table_specs_cost}

Building a sovereign model offers a government significant autonomy and the potential for greater long term value compared to buying a commercial service. 
However, the path to building is technically demanding and resource intensive. 
It is essential to highlight the specific challenges involved so that decision-makers can accurately assess whether they possess the necessary foundation to undertake such a project. 

Figure~\ref{fig:decision_points} summarizes the key decision points that arise throughout the lifecycle of building a large language model, and highlights a sequence of inter-dependent decision points spanning pre-building planning, training and evaluation, deployment and long-term iteration. 
These decision points reflect not only technical choices (e.g., model scope, data strategy, training stages), but also organizational, legal, economic and governance considerations that governments must address at different stages.

\begin{itemize}
    \item \textbf{Expertise}:
    The technical success of a sovereign build depends on a rare combination of expertise across multiple domains including data engineering, infrastructure systems, model training and red-teaming. 
    As established in 2025 labor reports, the global supply of these specialized professionals is exceptionally thin, often leading to ``talent wars'' where top tier engineers command compensation packages that rival professional athletes~\cite{recruitonomics2025}. 
    For any institution aiming to build LLMs, the challenge is not just the initial hire, but the retention of institutional memory: the departure of a single lead architect can delay a frontier model project by months or even years.
    Beyond internal technical expertise, a vibrant domestic open-source ecosystem is also an important resource. 
    A mature developer community acts as a force multiplier which provides a network that can rapidly stress test architectures, build derivative tools, and accelerate the localization of frontier technology.

    \item \textbf{Data}: 
    The effectiveness of any model is fundamentally limited by the quality and composition of its training data. 
    This process requires a massive effort to collect, clean, and annotate information while ensuring broad domain coverage. 
    Developers must also master a precise mixing recipe that balances diverse data sources to ensure the model gains general reasoning abilities without losing specialized knowledge. 
    As a reference, the Llama 3 technical report~\cite{dubey2024llama} describes a massive pre-training dataset of over 15 trillion tokens, where the data mixture was meticulously tuned to balance code, multilingual content, and high-quality reasoning data.
    Similarly, DeepSeek-V3~\cite{liu2024deepseek} details a strategy involving $14.8$ trillion tokens and a post-training phase that utilizes knowledge distillation to refine the "recipe" for specific task capabilities.
    Furthermore, data requirements shift throughout the training life-cycle; pre-training requires trillions of diverse tokens to build a knowledge base, while mid-training and post-training demand highly curated datasets for instruction following and safety alignment, a strategy emphasized in the Gemma 2~\cite{team2024gemma} report which prioritized data quality and distillation over sheer volume. 
    We provide detailed statistics on training data and token counts in Table~\ref{tab:specs_costs}, compiled from publicly available model reports.

    \item \textbf{Architectural Design Choices}:
    A sovereign build also requires governments to make a series of early architectural design choices that often require deep, hard-earned experience. 
    Decisions regarding model architecture, context length, tokenizer, and vocabulary size, as well as modality support, training techniques (e.g., full precision or mixed precision) and objective functions, all these design questions are rarely solved analytically and require extensive experience. 
    For a team without prior large model training experience, this architectual ``design tax'' can be as significant, and should therefore be treated as a first-order consideration in any decision to build. 
    We compile representative architectural specifications and training costs from several open-source LLM reports and summarize them in Table~\ref{tab:specs_costs}.

    \item \textbf{Training Compute Demands}:
    The physical infrastructure required for training is substantial and expensive. 
    For reference, training a mid-scale model with $70$ billion parameters, such as Llama 3 70B, can require approximately $6.4$ million GPU hours~\cite{MetaLlama3ModelCard}, while the subsequent Llama-3.3 70B utilized roughly $7.0$ million GPU hours for its pre-training phase~\cite{NVIDIA_NIM_API_Llama33}.
    The architectural choice and training technique adoption can make a difference: as reported in DeepSeek-V3 report~\cite{liu2024deepseek}, training a $671$B DeepSeekMoE architecture with the support of FP8 training requires only $2.8$ million H800 GPU hours for its full training.
    Additionally, managing these runs requires robust training frameworks capable of handling checkpointing and failure recovery. 
    At this scale, even a brief hardware interruption can result in million dollar losses in wasted compute time. 
    Estimating the training cost must account for unexpected failures and interruptions. For example, Meta has reported in~\cite{dubey2024llama} that during the 54-day pretraining of Llama 3 $405$B model, more than half of the $419$ unexpected interruptions were caused by failures related to GPUs or their onboard HBM3 memory.
    Additionally, many of the most effective training techniques remain proprietary or poorly documented, which increases the risks for teams without extensive experience.
    
    \item \textbf{Evaluation}: 
    Evaluation measures the operational utility of a model across various dimensions, including instruction following ability~\cite{jing2023followeval, wen2024benchmarking, wangpandalm}, domain-specific proficiency (e.g. coding~\cite{jimenez2024swe} and math~\cite{lightman2023lets, amini2019mathqa}), context window reliability~\cite{chen2025longleader}, efficiency and so on. 
    Beyond relying on generic leaderboards, governments developing sovereign models must design evaluation frameworks that reflect national priorities rather than purely academic or commercial benchmarks. 
    In practice, governments often have to decide whether to build their own benchmarks: an example was the case in SEA-LION~\cite{ng2025sealion} where there were no high-quality evalution datasets for Lao, Khmer, or regional multi-turn dialogues in Malay and Tagalog, requiring the creation of new benchmarks such as SEA-HELM~\cite{susanto2025sea}.
    Constructing sovereign benchmarks brings its own challenges. 
    Localizing test content requires extensive human expertise and cannot be easily automated or synthetically generated at scale - and the use of LLM-as-a-judge methods introduces a ``chicken-and-egg'' problem.

    \item \textbf{Alignment, Regulation and Red-Teaming}:
    Once a model is trained, it must undergo rigorous alignment to ensure its behavior is safe, lawful and consistent with national or regional norms.  
    This involves constraining the model to reduce the likelihood of generating harmful, misleading, or non-compliant outputs, often requiring constructing specific datasets and evaluation sets, which are time-consuming and manpower-intensive.
    In addition, robust red-teaming is essential: agencies should proactively stress-test LLMs for adversarial prompting~\cite{liu2023prompt, liu2024formalizing}, jailbreaking~\cite{zou2023universal, yi2024jailbreak} and any other policy-violating behaviors~\cite{amine2025prompt}. 
    Systematic red-teaming has become a standard pre-deployment practice in the industry: leading model providers such as OpenAI~\cite{ahmad2025openai, beutel2024diverse}, Anthropic~\cite{anthropic2024redteaming} and Google~\cite{gemini2025safeguards} routinely conduct extensive adversarial testing to ensure safety before releasing large models.
    For government agencies, compliance with sector specific regulations and auditing standards is non-negotiable and a continuous commitment.
    
    \item \textbf{Inference and Deployment}:
    The final challenge lies in the production environment where scalability and latency are critical. 
    LLMs are expected to process high volumes of concurrent requests while maintaining consistent response speeds.
    Maintaining operational viability involves balancing computational throughput with cost efficiency to prevent infrastructure overhead from exceeding sustainable levels.
    Deployment choices also shape governance: production systems need robust safeguards against prompt injection, data exfiltration, and misuse, as well as auditable logging and access controls that meet public-sector requirements.
    In practice, governments may rely on mature hosting and MLOps platforms to reduce operational burden. For example, managed model hubs and inference services (e.g., Hugging Face), hyperscaler cloud platforms and model marketplaces (e.g., AWS, Azure, Google Cloud), or enterprise AI platforms (e.g., IBM); or adopt hybrid approaches combining on-premise/sovereign-cloud hosting with vendor tools. 
    Ultimately, the practical impact of an AI system depends on a structured roadmap for deployment and tool orchestration that ensures reliability in real-world daily public service delivery.
\end{itemize}

%% file: figures/table_specs_cost.tex
\begin{table}[t!]
\caption{\textbf{Model specifications and associated training costs.} ``--'' indicates values not reported. For each model family, we list the flagship configuration, which is typically the largest model. Additional configurations and cost details for smaller variants can be found in the cited sources. Please note that reported training cost in these literature often refer only to the ``final run'', rather than the full end-to-end development cost.}
\label{tab:specs_costs}
\vspace{0.5em}
\resizebox{\textwidth}{!}{
\begin{tabular}{lccccccccc}
\toprule
& \multicolumn{5}{c}{\textbf{Model specifications}} 
& \multicolumn{4}{c}{\textbf{Training effort \& cost}} \\
\cmidrule(lr){2-6} \cmidrule(lr){7-10}
\textbf{Model} &
\textbf{Params} &
\textbf{Arch} &
\textbf{\makecell{Context\\Length}} &
\textbf{\makecell{Tokenizer}} &
\textbf{\makecell{Vocabulary\\Size}} &
\textbf{Trained Tokens} &
\textbf{\#Chips} &
\textbf{Training Hours} &
\textbf{Cost (USD)} \\
\midrule
\rowcolor{gray!60!red!10!white}DeepSeek V3~\cite{liu2024deepseek} &
\makecell{671B-A37B\\(671B total,\\37B activated)} &
MoE &
128K &
\makecell{BBPE (byte-level\\byte-pair encoding)\\\cite{shibata1999byte, brown2020language}} &
129{,}280 &
14.8T &
2048 H800 GPUs &
\makecell{2.788M\\H800 GPU hours} &
\$5.576M \\
\addlinespace
gpt-oss~\cite{agarwal2025gpt} &
120B-A5B &
MoE &
131{,}072 &
o200k\_harmony~\cite{tiktoken} &
201{,}088 &
-- &
-- &
\makecell{2.1M\\H100 GPU hours} &
-- \\
\addlinespace
\rowcolor{gray!60!red!10!white}Kimi K2~\cite{team2025kimi} &
1T-A32B &
MoE &
128K &
tokenizer &
160K &
\makecell{Pretraining: 15.5T\\Long-context: 460B} &
-- &
-- &
-- \\
\addlinespace
Qwen3~\cite{yang2025qwen3} &
235B-A22B &
MoE &
128K &
BBPE &
151{,}669 &
36T &
-- &
-- &
-- \\
\addlinespace
\rowcolor{gray!60!red!10!white} Qwen3-VL~\cite{Bai2025Qwen3VLTR} &
235B-A22B &
\makecell{LLM: MoE\\Vision: SigLIP-2~\cite{tschannen2025siglip}} &
256K &
BBPE &
151{,}669 &
$\sim$2.1T &
-- &
-- &
-- \\
\addlinespace
 Step-3~\cite{step3model, step3system} &
321B-38B &
\makecell{LLM: MoE\\Vision:Eva-CLIP 5B~\cite{sun2023eva}} &
65536 &
BBPE &
-- &
\makecell{$>20$T for language,\\$4$T multimodal} &
-- &
-- &
\makecell{Per token training cost\\ $\$ 0.213$ (Fig.4 of \cite{step3system})} \\
\addlinespace
\rowcolor{gray!60!red!10!white}Step3-VL-10B~\cite{huang2026step3} &
\makecell{LLM: 8B\\Vision: 1.8B} &
\makecell{LLM: dense Transformer\\Vision: Perception Encoder~\cite{bolya2025perception}} &
128K &
Qwen3~\cite{yang2025qwen3} &
151{,}669 &
\makecell{Pretraining: $1.2$T\\SFT: $226$B\\RL:$>1400$ iters} &
-- &
-- &
-- \\
\addlinespace
Gemma 3~\cite{team2025gemma} &
\makecell{LLM: 27B\\Vision: 400M} & 
\makecell{LLM: dense Transformer\\Vision: SigLIP~\cite{zhai2023sigmoid}} &
128K &
SentencePiece~\cite{kudo2018sentencepiece} &
262{,}000 &
14T &
6144 TPUv5p &
-- &
-- \\
\addlinespace
\rowcolor{gray!60!red!10!white}Llama 3~\cite{dubey2024llama} &
\makecell{LLM: 405B\\Vision: 630M\\Speech: 1B} &
\makecell{LLM: Dense Transformer\\Vision: ViT-H/14~\cite{dosovitskiy2021image}\\Speech: Conformer~\cite{gulati2020conformer}} &
128K &
tiktoken BPE~\cite{tiktoken} &
128{,}000 &
15.6T &
16K H100 GPUs &
-- &
-- \\
\addlinespace
Llama-Nemotron~\cite{bercovich2025llama} &
405B &
\makecell{improved arch by NAS\\and FFN Fusion based on~\cite{dubey2024llama}} &
128K &
tiktoken BPE~\cite{tiktoken} &
128{,}000 &
\makecell{Knowledge distillation: 65B\\Continued pretraining: 88B} &
-- &
\makecell{RL stage: 140k\\H100 GPU hours} &
-- \\
\addlinespace
\rowcolor{gray!60!red!10!white}Mistral-Large-3~\cite{mistral_large_3} &
\makecell{LLM:675B-A41B\\Vision:2.5B} &
MoE &
256K &
-- &
-- &
-- &
3000 H200 GPUs &
-- &
-- \\
\addlinespace
ERNIE 4.5~\cite{ernie2025technicalreport} &
300B-A47B &
\makecell{Heterogeneous MoE\\Vision: ViT} &
131{,}072 &
-- &
--
&
-- &
2016 H800 GPUs &
-- &
-- \\
\addlinespace
\rowcolor{gray!60!red!10!white}GLM-4.5~\cite{zeng2025glm} &
355B-A32B &
MoE &
128K &
-- &
-- &
23T &
-- &
-- &
-- \\
\addlinespace
MiniMax-Text-01~\cite{li2025minimax} &
456B-A46B &
MoE &
1 million &
BBPE &
200K &
Pretraining: $11.4$T &
$1500\sim2500$ H800 GPUs &
-- &
-- \\
\addlinespace
\rowcolor{gray!60!red!10!white}MiniMax-VL-01~\cite{li2025minimax} &
\makecell{LLM: 456B-A46B\\Vision: 303M} &
\makecell{LLM: MoE\\Vision: ViT}     &
1 million &
BBPE &
200K &
\makecell{Pretraining: $11.4$T\\Multimodal: $512$B} &
$1500\sim2500$ H800 GPUs &
-- &
-- \\
\addlinespace
MiniMax-M1~\cite{chen2025minimax} &
456B-A46B &
MoE &
1 million &
BBPE &
200K &
\makecell{Based on~\cite{li2025minimax},\\Continue pretraining: 7.5T\\RL: 3 weeks} &
512 H800 GPUs &
\makecell{RL stage: $\sim258048$\\H800 GPU hours} &
\makecell{RL stage: $\$0.53$M}\\
\addlinespace
\rowcolor{gray!60!red!10!white}Aria~\cite{li2024aria} &
\makecell{LLM:25B-A4B\\Vision:438M} &
\makecell{LLM:MoE\\Vision:SigLIP-SO400M~\cite{zhai2023sigmoid}} &
64K &
-- &
-- &
\makecell{Pretraining:6.4T\\Multimodal:400B} &
-- &
-- &
-- \\
\addlinespace
Phi-4~\cite{abdin2024phi} &
14B &
dense Transformer &
16K &
tiktoken~\cite{tiktoken} &
100{,}352 &
\makecell{Pretraining:10T\\Midtraining:250B\\SFT:8B;\ \  DPO:850k pairs} &
-- &
-- &
-- \\
\addlinespace
\rowcolor{gray!60!red!10!white}Olmo 3.1~\cite{olmo2025olmo} &
32B &
dense Transformer &
65{,}536 &
 cl100k~\cite{gpt3_5turbo} &
 50{,}304 &
 6.1T &
 1024 H100 GPUs &
 \makecell{$\sim$ 1.376M\\ H100 GPU hours} &
 $\sim\$$2.75M \\
\addlinespace
Apertus-70B~\cite{hernandez2025apertus} &
70B &
dense Transformer &
65{,}536 &
v3 tekken~\cite{mistral_nemo} &
131{,}072 &
$\sim$15T &
4096 H200 GPUs &
\makecell{$\sim$6M\\ H200 GPU hours} &
-- \\
\addlinespace
\rowcolor{gray!60!red!10!white}Falcon 3~\cite{Falcon3} &
10B &
dense Transformer &
32K &
-- &
131K &
\makecell{Pretraining: 14T\\Continued pretraining:2T} &
1024 H100 GPUs &
-- &
-- \\
\addlinespace
Falcon-H1~\cite{zuo2025falcon} &
34B &
Hybrid Mamba-Transformer &
16K &
BPE &
261K &
\makecell{Total:$\sim18$T\\Pretrain(multilingual):3T\\} &
4096 H100 GPUs &
-- &
-- \\
\addlinespace
\rowcolor{gray!60!red!10!white}Falcon-H1-Arabic~\cite{Falcon-H1-Arabic-2026} &
34B &
Hybrid Mamba-Transformer~\cite{zuo2025falcon} &
256K &
BPE &
261{,}120 &
\makecell{Pretrain(Arabic): 160.9B\\Pretrain(English): 155.9B} &
-- &
-- &
-- \\
\addlinespace
Llama-SEA-LION-v3-70B-IT~\cite{sea-lion-v3} &
70B &
dense Transformer &
128K &
tiktoken BPE~\cite{tiktoken} &
128{,}000 &
\makecell{Continue pretraining:200B\\Instruction tuning on \\(English): 12.3M pairs; (SE Asian): 4.5M pairs} &
\makecell{64 H200 GPUs and\\128 H100 GPUs} &
\makecell{12800 H200 GPU hours and\\63360 H100 GPU hours} &
-- \\
\addlinespace
\rowcolor{gray!60!red!10!white}Gemma-SEA-LION-9B-IT~\cite{ng2025sealion} &
9B &
dense Transformer &
8192 &
SentencePiece &
256{,}000 &
\makecell{$\sim$200B for\\ continued pretraining} &
8 H100 GPUs&
\makecell{1350\\H100 GPU hours} &
-- \\
\bottomrule
\end{tabular}
}
\end{table}


%% file: sections/framework.tex
\section{Strategic Evaluation Framework}
\label{sec:strategic_framework}
\begin{figure}[h!]
  \centering
  \includegraphics[width=0.85\linewidth]{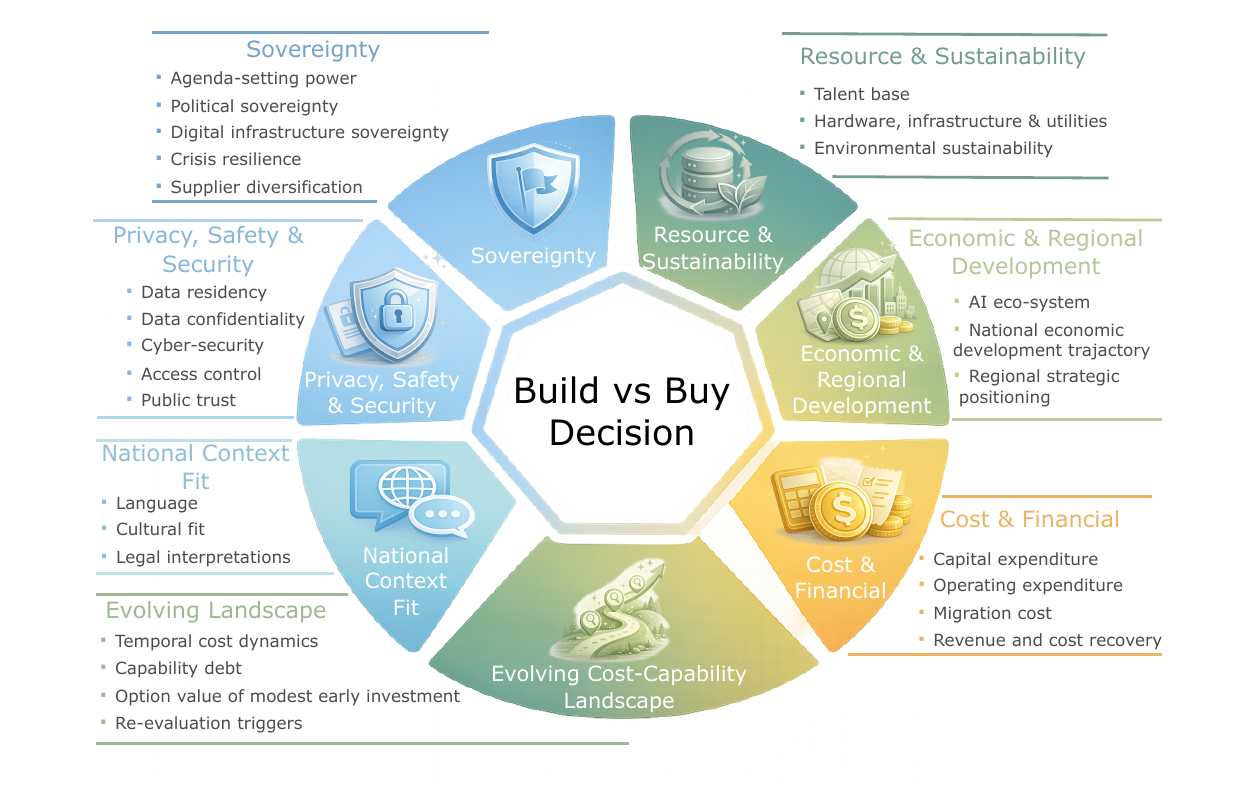}
  \caption{\textbf{Strategic evaluation framework for government LLM build-vs-buy decisions}. }
  \label{fig:framework}
\end{figure}

The choice between building and buying must be weighed against a set of strategic dimensions that impact the long term health of public institutions.
As illustrated in Figure~\ref{fig:framework}, government decisions on whether to build or buy LLMs must be evaluated across multiple strategic dimensions, spanning sovereignty, security, resource capability and sustainability, economic and regional development, cost and financial, national context fit and evolving cost-capability landscape.
Decision-makers could consider the pros and cons of each option from the following perspectives:
\vspace{-0.5em}

\subsection{Sovereignty Perspective}

This dimension addresses how a deployment choice affects national autonomy and strategic dependence. 
Governments must evaluate several critical layers of control:

\begin{itemize}
    \item \textbf{Agenda-setting Power over Norms, Values, and Information}:
    Large language models serve as more than simple search engines because they control the narrative layer of citizen interaction. 
    These systems decide how issues are framed and what is treated as a reasonable or factual conclusion. 
    By relying on a third-party model, a government may inadvertently cede control over how sensitive topics like history or national identity are handled. Owning the model allows an institution to ensure that the logic and values embedded in the system remain consistent with local culture and national norms.
    
    \item \textbf{Political Sovereignty}: 
    Model policies can directly influence political discourse and democratic processes. 
    For instance, during the 2024 US election cycle, OpenAI implemented several specific restrictions to manage political content, including elevating authoritative sources of information, preventing deepfakes and disrupting threat actors~\cite{OpenAI2024Election}. 
    A sovereign build allows a government to set its own rules for political neutrality and discourse within its jurisdiction.

    \item \textbf{Digital Infrastructure / Asset Sovereignty}: 
    In the coming years, these models will function as essential public infrastructure similar to identity systems or healthcare platforms. 
    Asset sovereignty concerns the ability to maintain long term control over these foundations. 
    This includes the power to set usage limits, modify alignment rules, or audit operational logs and data. 
    A buy strategy may leave a government vulnerable to a vendor choosing to change pricing structures, alter system performance, or even suspend service without warning.
    
    \item \textbf{Crisis Resilience}:
    During emergencies such as cyber-attacks or major disasters, the continuity of public services is paramount. 
    Governments must consider whether they can guarantee the integrity of an automated service if the primary provider is located outside their borders or suffers a systemic failure. 
    A build strategy provides a higher degree of resilience by ensuring that the technology can operate within isolated or high security environments during a national crisis.

    \item {\textbf{Supplier Diversification}: Beyond ownership of a single model, sovereignty also depends on avoiding concentration risks across the AI stack. 
Governments should consider diversification across software providers, hardware platforms, critical talent, and energy supply to reduce lock-in risks and single points of failure. 
Such diversification strengthens operational resilience under both buy and build strategies, particularly during geopolitical shocks or supply disruptions.}
\end{itemize}

\subsection{Privacy, Safety and Security Perspective}

This dimension evaluates the protection of sensitive government and citizen information throughout the model life-cycle. 
Decisions must account for all layers of data protection to maintain institutional integrity.

\begin{itemize}
    \item \textbf{Data Residency}:
    Governments must determine exactly where data is stored and processed. 
    Relying on external cloud providers often involves information crossing national borders, which may subject citizen data to foreign laws and surveillance. 
    Localized or on premises deployment ensures that sensitive records remain within the legal and physical jurisdiction of the state, eliminating the risk of extraterritorial access.
    
    \item \textbf{Data Confidentiality}: 
    Beyond basic data separation, governments must guard against advanced technical threats such as model inversion~\cite{sivashanmugam2025model} or prompt inversion attacks~\cite{morris2024language, zhang2024extracting}. 
    These vulnerabilities could allow adversaries to reconstruct sensitive training data or private user prompts from the model outputs. 
    A build strategy allows for deeper implementation of privacy enhancing technologies to mitigate these specific risks.

    \item \textbf{Cyber-security}:
    LLMs increasingly possess strong code-writing and agentic capabilities~\cite{xu2024large, hou2024large, lupinacci2025dark, kozak2025developer}, which introduces both risks and opportunities for national cyber-security. 
    On the risk side, LLMs may be misused to generate malware, exploit code, or automate cyber-attacks if safeguards are insufficient. 
    On the opportunity side, the same capabilities can be leveraged for defensive purposes, such as vulnerability analysis, code auditing, incident response, and cyber-threat intelligence. 
    LLMs should be treated as cyber-security infrastructure, where deployment choices determine both misuse risks and defensive capabilities.

    \item \textbf{Access Control}: 
    Managing who can view interaction logs, prompts, and system outputs is vital for preventing internal data breaches. A robust framework requires granular permissions and clear auditing trails. While commercial services offer some controls, a private deployment grants the government total authority over the entire stack, ensuring that no third party can access the data flowing through the system.
    
    \item \textbf{Public Trust}: 
    The legitimacy of government AI depends on public confidence in how these systems manage and protect their information. 
    Public confidence is typically higher when data is managed by domestic institutions rather than foreign corporations. 
    Building trust requires active consultation with civil society to ensure that the deployment aligns with community expectations and ethical standards.
\end{itemize}

\subsection{Cost and Financial Perspective}

From an economic perspective, governments should evaluate the buy–build decision considering upfront capital requirements, recurring operating expenses, switching costs, and the potential for cost recovery over time.

\begin{itemize}
    \item \textbf{Capital Expenditure (CAPEX)}: 
    A build strategy requires significant upfront investment in physical hardware, such as specialized GPU clusters and high performance networking. 
    It also involves the cost of acquiring massive datasets and hiring a team of elite technical experts. 
    These projects carry higher trial-and-error risks where a failure can lead to substantial financial loss. 
    In contrast, a buy strategy has much lower initial capital requirements, with costs mainly focused on security reviews, network integration and staff training for third-party interfaces.

    \item \textbf{Operating Expenditure (OPEX)}: 
    For a built model, operating costs are relatively stable but significant, covering energy consumption, hardware maintenance, and permanent staff salaries. 
    However, a volatile and easily overlooked component of OPEX is the cumulative expenditure on experimental overhead - before achieving a final successful training run, the team must fund ablation studies and hyperparameter tuning sessions to identify the optimal model configuration.
    According to the Frontier Model Training Costs report by Epoch AI~\cite{cottier2024rising}, the total compute for model development is $1.2$x to $4$x larger than the final training run.
    For a bought model, costs scale directly with usage volume. 
    While this provides flexibility, it also creates budget uncertainty because vendors can change pricing models or discontinue teaser rates once a department is integrated into their ecosystem. 
    Over time, high volume government use can make a subscription more expensive than the maintenance of an internal system.
    
    \item \textbf{Migration Costs}: 
    The transition between model architectures or service providers entails fiscal and technical expenditures.
    A government must consider the technical debt and integration costs involved in moving data and workflows to a new system. 
    Building a private model reduces the risk of vendor lock in, whereas a buy strategy might make an agency dependent on a single provider’s proprietary architecture.

    \item \textbf{Revenue and Cost Recovery}: 
    Building an internal model allows the government to treat the system as a national asset. 
    The government can recover costs by charging commercial entities or high volume public users for access, effectively reinvesting that revenue into the domestic economy. 
    When buying, fees flow primarily to external vendors.
    
\end{itemize}


\subsection{Economic and Regional Development Perspectives}

Beyond immediate costs and technical feasibility, the decision to build or buy LLM capabilities can shape a country’s long-term industrial structure, innovation ecosystem, and regional strategic position.

\begin{itemize}
\item \textbf{AI Eco-system}: 
    The decision can significantly influence the national technological landscape by determining how universities, labs and startups evolve.
    A build strategy acts as a direct catalyst for the domestic tech sector, creating high value research roles and securing sovereign intellectual property.
    Conversely, a buy strategy stimulates economic growth by providing a high performance foundation for a secondary innovation layer, allowing local developers to focus on specialized vertical applications.
    Ultimately, the choice involves a trade-off between developing long-term foundational independence and fostering rapid application adoption across the wider business community.

    \item \textbf{National Economic Development Trajectory}:
    The buy–build decision should also be viewed in light of a country’s long-term economic development strategy. 
    Governments pursuing a transition toward a technology- and knowledge-driven economy may place higher strategic value on building domestic LLM capabilities, as such investments can stimulate high-skilled employment, upstream infrastructure development, and downstream innovation across sectors.
    By contrast, countries whose economic structures are expected to remain primarily agriculture-, resource-, or extraction-based may rationally prioritize access to AI capabilities as users rather than builders, focusing on cost efficiency and rapid deployment over domestic model development. 
    In this sense, the decision to build or buy is not purely technical or financial, but closely tied to national industrial policy and growth objectives.

    \item \textbf{Regional Strategic Positioning}:
    Even if building domestic LLMs is unlikely to place a country among global AI leaders, it can still confer meaningful regional advantages relative to neighboring or peer countries. 
    Indigenous AI capabilities may strengthen a country’s bargaining position in regional cooperation, reduce reliance on and increase bargaining power with external technology providers, and allow it to set standards within shared markets or institutions.
    In some cases, serving as an early regional adopter or developer of government-grade AI systems can attract talent, partnerships, and investment, reinforcing the country’s strategic relevance within its region.
\end{itemize}

\subsection{Resource Capability and Sustainability Perspective}
\vspace{-0.25em}

This dimension assesses whether a nation possesses or intends to develop the technical foundations required to train and operate advanced LLMs. 
Decisions in this area determine the long-term feasibility of  sovereign build strategy.

\begin{itemize}
    \item \textbf{Talent Base}: 
    A successful build strategy requires a strong team of experts across hardware infrastructure, data engineering, algorithm, security and governance. 
    The central challenge lies in the government's ability to compete with the private tech companies to attract and keep these professionals, who can often be recruited by tech firms with high-paying job offers. It is obvious that having a strong national research-intensive universities is critical for attraction, retention and sustained production of talent. Some countries rely on international talent while that might not be an option for other countries.

    \item \textbf{Hardware, Infrastructure and Utilities}:
    A sovereign build strategy requires a robust physical foundation, including high-performance GPUs, low-latency networking and a massive, reliable supply of electricity and cooling.
    A build strategy is only viable if the country has reliable access to these physical assets or maintains a strategic reserve to withstand global market volatility.
    

    \item \textbf{Environmental Sustainability}:  
    Large-scale language models can be energy-intensive across training, deployment, and ongoing operations, making environmental sustainability a critical consideration.
    Building and operating an in-house model entails a non-trivial electricity demand and associated carbon footprint, which may conflict with national climate commitments unless paired with low-carbon power procurement and efficiency measures. 
    For instance, the training of the $176$B BLOOM model consumed $433,196$ kWh total energy and was estimated to emit about $24.69$ tCO$_2$e~\cite{luccioni2023estimating}\footnote{tCO$_2$e: (tonnes of carbon dioxide equivalent), a standard measure that expresses the total climate impact of greenhouse gas emissions on a common CO$_2$ basis.}.
    Importantly, environmental impacts do not end at training: the inference stage can accumulate substantial footprint at scale.
    Mistral’s lifecycle assessment~\cite{Mistral2025report} reports that by January 2025, Mistral Large 2’s training plus 18 months of usage amounted to $20.4$ ktCO$_2$e, $281,000$ m$^3$ of water consumption, and $660$ kg Sb eq\footnote{kg Sb eq: (Kilograms of antimony equivalent), a life-cycle assessment metric indicating depletion of scarce mineral resources, normalized to antimony as a reference.} of resource depletion. 
    On average, a single $400$-token response (excluding end-user devices) contributes an estimated $1.14$ gCO$_2$e, $45$ mL of water, and $0.16$ mg Sb eq~\cite{Mistral2025report}.
    Under a buying strategy, these impacts are often operationally externalized to the provider’s infrastructure, but governments may still face indirect accountability through procurement standards, public reporting expectations, and alignment with national sustainability targets.
\end{itemize}

\subsection{National Context Fit Perspective}
\vspace{-0.25em}

This dimension evaluates how effectively a model aligns with the specific linguistic, legal, and social realities of a nation. 
For government agencies, a model that lacks local context can become a liability, leading to misinterpretations of policy or alienating the citizens it is meant to serve.
\vspace{-0.5em}

\begin{itemize}
    \item \textbf{Language}: 
    Language coverage is one of the core determinants of the decision especially in multilingual societies. 
    If the primary language is already well supported by global tech companies, buying can deliver strong performance with minimal upfront effort.
    In contrast, for the purpose of serving minority languages and regional dialects where commercial models may struggle with, the building strategy allows governments to prioritize these low-resource languages using their own specialized datasets.
    Building can therefore improve inclusive access to public services and help preserve linguistic identity, but it requires sustained effort in data collection, evaluation, and ongoing maintenance.

    \item \textbf{Cultural Fit}:
    The ability to handle historical references, political context, and local social norms is vital for maintaining public trust. 
    Cultural fit involves more than just language; it requires the model to respect religious beliefs, understand local taboos, know local cuisines and follow regional customs. 
    Furthermore, many countries require fine-grained control to account for different laws or values across various states and provinces. 
    In some countries, rich cultural traditions are gradually eroding as younger generations increasingly adopt global cultural norms.
    Governments may view culturally aligned LLMs as a tool to support cultural continuity, particularly in education, public communication and youth engagement.
    A locally developed or highly customized model can be programmed with these specific sensitivities in mind, ensuring that its outputs are not only accurate but also culturally appropriate and socially responsible.

    \item \textbf{Legal Interpretations}: Can LLMs embed national legal frameworks, case law standards, and procedural rules into its parameters? 
    An LLM should be able to embed national statutes, case law standards, and administrative rules directly into its parameters or retrieval systems. 
    In practice, general-purpose models could default to assumptions from dominant jurisdictions from foreign laws, producing answers that may be misleading, inappropriate, or even unlawful in other countries.
    When pursuing a buying strategy, governments may therefore need to explicitly negotiate legal disclaimers and domain constraints with model providers to mitigate these risks. 
    On the other hand, building carries risk: without sufficient data and building capability, such systems can produce misleading, hallucinated outputs or even illegal advice~\cite{mamdani2026shut}, wasting resources and undermining trust. 
    A recent example is New York City's internally developed legal chatbot, which was shut down after producing unreliable advice despite costing over $\$600{,}000$ to build~\cite{mamdani2026shut}.
    In such cases, combining reliable priopriety models with controlled retrieval (e.g., RAG) may be the safer choice.
    
\end{itemize}

\subsection{Evolving Cost-Capability Landscape Perspective}
\label{sec:evolving_landscape}

This dimension emphasizes that the buy-build decision is not a one-time choice.
The optimal strategy can shift over time as technology diffuses, costs evolve, and domestic capability accumulates. 
Governments should therefore treat the decision as a dynamic planning problem  rather than as a point-in-time procurement comparison.

\begin{itemize}
    \item \textbf{Temporal Cost Dynamics}: The buy-build decision can shift substantially over time as the underlying cost--capability frontier moves. 
    Training costs may decline rapidly; for example, DeepSeek V3 reports a training cost of roughly \$5.6M, an order-of-magnitude reduction relative to earlier frontier-scale efforts. 
    In parallel, the open-source ecosystem continues to mature, with open models increasingly approaching frontier performance. 
    Compute hardware is also becoming more widely available, while local talent pools can deepen through knowledge transfer and the growing transparency of technical reports and academic publications.
    Together, these potential trends can reduce the cost and raise the feasibility of building, such that a government may rationally choose \emph{buy} now and \emph{build} in the next few years, not because its preferences changed, but because the underlying cost--capability landscape shifted.

    \item \textbf{Capability Debt}: LLM development capability is not a function of capital investment alone; it is cumulative, depending on hands-on training experience, institutional know-how, data pipelines and partnerships, and operational lessons learned through iteration and failure. 
    Delaying capability building can therefore create \emph{capability debt}, raising the cost and difficulty of catching up even as absolute compute costs decline.

    \item \textbf{Option Value of Modest Early Investment}: Modest early investments such as small-scale training pilots, sovereign benchmarks, and infrastructure prototyping can preserve future build options without immediate commitment to full-scale training. 
    These efforts accumulate organizational know-how, reduce future transition costs, and enable faster scaling if priorities or constraints change. 
    This option value is often omitted in static cost comparisons but can be strategically material.

    \item \textbf{Re-evaluation Triggers}: 
    Governments can make regular re-evaluation by setting clear triggers, such as widening capability gaps, major improvements in open models, vendor price or policy changes, new regulations, or clear progress in domestic talent and infrastructure. 
    Defined triggers and review cycles ensure the strategy keeps pace with the technology landscape rather than changing only in response to crises.
\end{itemize}

Taken together, these perspectives highlight that the build–buy decision for government LLM deployment is inherently multi-dimensional and cannot be reduced to a single technical or budgetary criterion.
Choices that optimize short-term cost efficiency may introduce long-term dependencies in sovereignty, resilience, or ecosystem development, while investments in sovereign capability demand sustained commitments in resources, governance, and sustainability.
The appropriate balance depends on national priorities, institutional capacity, and risk tolerance, as well as how governments weigh autonomy, public trust, and long-term economic positioning against speed of deployment and operational simplicity.
Rather than prescribing a universal answer, this framework is intended to help decision-makers surface trade-offs explicitly and align LLM deployment strategies with broader national objectives.

%% file: sections/opinionated.tex
\section{Insights from Practice: Lessons from Building Language Models}

This section distills practical experience-based insights from the paper's authors, including practitioners who have led or contributed to national-level public-sector LLM efforts, as well as industry experts with deep experience in large-scale model development. 
Presented as curated Q\&As, they reflect practical lessons and decision trade-offs encountered during the development of Singapore's SEA-LION~\cite{ng2025sealion} and Switzerland's Apertus~\cite{hernandez2025apertus}, two prominent examples of government-backed sovereign LLM efforts, where members of the author team directly led or co-led the model development.

We particularly wish to avoid a misleading impression that a country builds a single sovereign LLM and mandates its use across all domains and applications. 
In practice, national AI strategies are far more pluralistic, and in neither country are these models mandated for universal use; instead, they coexist with commercial and open-source models which are used in many non-sensitive or general-purpose settings. These cases illustrated how sovereign models can complement the broad model ecosystem, serving specific strategic, linguistic, or public-interest needs.

\vspace{1em}
\begin{tcolorbox}[question]
\textit{Which factors were decisive in the choice of building and the selection of technical route?}
\end{tcolorbox}
\vspace{1em}

\subsubsection*{Switzerland: Apertus}

For the Apertus project, the decision to build was triggered by the launch of Switzerland's Alps supercomputing cluster, equipped with $11{,}000$ GH200 GPUs. The availability of large-scale, publicly funded infrastructure created a timely opportunity to evaluate whether such systems could support frontier-scale AI workloads, directly catalyzing this project.

Within Switzerland's academic- and public-sector led environment, building a sovereign LLM was motivated by several strategic considerations. 
\textbf{AI sovereignty and capability} were central: maintaining control over AI systems was viewed as increasingly important given their future role on technological innovation.
\textbf{Responsibility and democratization} were equally influential. The team sought to demonstrate that large-scale models can be developed responsibly and in the public interest. By making Apertus fully open and modular, the project aimed to lower barriers for regions and institutions that lack resources to build full AI pipelines, enabling them to adopt or extend selected components. Together, these factors led Apertus to pursue a build-oriented, open and standards-setting technical approach aligned with long-term public values.

The decision to pretrain from scratch was also driven by the commitment to establishing a responsible foundation for AI development. Much of today's AI ecosystem follows a ``rich get richer'' dynamic where large companies aggregate vast amounts of data with limited consideration for data ownership. Through Apertus, the goal was to challenge this trajectory by respecting data rights, public values and social responsibility, while still delivering practical utility for real-world applications.

\subsubsection*{Singapore: SEA-LION}
For SEA-LION models, the decision to build was triggered at the national level in late 2022, shortly around the time of ChatGPT's release. The main motivation was capability building – Singapore sought to develop in-house expertise in training LLMs and to secure a strategic position in generative AI. Security concerns around using public APIs for sensitive applications were a secondary factor.

At the time of building SEA-LION v1, restrictive licenses of available open models (e.g., LLaMA-1) further reinforced the decision to pretrain from scratch.  In particular, LLaMA-1 did not permit commercial use, which conflicted with our goal of maximizing social impact by enabling wide adoption including enterprises. As a result, building on top of LLaMA-1 was not viable, and pretraining a commercially usable model from scratch become a necessary choice.

\begin{tcolorbox}[question]
\textit{What triggered the building of later versions (v2 - v4) of SEA-LION series?}
\end{tcolorbox}

SEA-LION v1 was a proof-of-concept. By pretraining small models from scratch (3B and 7B), the team aimed to acquire hands-on capability in building LLMs end to end. After its release in late 2023, the next trigger was the need to align with scaling law and to pursue higher capability.

By the time of SEA-LION v2, the landscape had shifted in two important ways. (1) First, more open models with permissive licenses became available, enabling broader downstream use. However, many of these base models lacked sufficient pretraining coverage for under-represented Southeast Asian languages (e.g., Khmer, Burmese, Lao), which remained a core requirement. 
(2) Second, frontier models were scaling rapidly in both parameters and training data volume, which makes full pretraining impractical for the government team constrained by legally usable data.
As a result, SEA-LION v2 and v3 adopted a hybrid strategy: leveraging open pretrained models while performing continual pretraining (CPT) and post-training on legally sourced, region-specific data.
This approach was first explored in v2 on an 8B model, and then scaled up to 70B in v3 .

For SEA-LION v4, we got practical feedback that indicated smaller models are more preferable in Southeast Asia due to deployment and inference cost constraints. 
At the same time, the team recognized that SEA-LION did not need to compete directly with frontier-scale models, given the continued growth of the open-LLM ecosystem.
Therefore, v4 deliberately scaled down from 70B to 27B, prioritizing efficiency while maintaining strong performance.

\vspace{1em}
\begin{tcolorbox}[question]
\textit{Which costs in large-scale model training are often underestimated before building begins?}
\end{tcolorbox}
\vspace{1em}

Many teams may underestimate the importance of training infrastructure, especially for large-scale pretraining (e.g., models exceeding hundreds of billions of parameters). 
While hardware can be purchased, scalable training software infrastructure, especially for frontier-scale models, cannot be easily acquired off the shelf. 
Open-source training stacks may suffice for smaller models, but they lag significantly behind commercial systems when dealing with frontier-scale pretraining workloads. 
A well-designed infrastructure can easily double training efficiency in both time and cost. 
Optimizing training infrastructure plays a critical role in reducing LLM training costs, as evidenced by documented examples such as DeepSeek-V3~\cite{liu2024deepseek}, Step-3~\cite{step3system} and so on.
Talent with experience in large-scale pretraining, low-level kernel optimization and systems such as Megatron is extremely scarce, making infrastructure development both technically and organizationally challenging.

\vspace{1em}
\begin{tcolorbox}[question]
\textit{How should governments decide the budget for building an LLM from scratch?}
\end{tcolorbox}
\vspace{1em}

One practical reference point for budget estimation is the market valuation of companies that have demonstrated the capability to build LLMs at the desired level. 
While such valuations reflect far more than the LLM itself, they nonetheless provide an order-of-magnitude signal of the resources required to assemble a team capable of building competitive LLMs. 

Importantly, we want to highlight that the budget is highly context-dependent: the required investment varies significantly based on existing assets, such as access to compute infrastructure, availability and cost of experienced talents, prior model development experience, and data resource already in place.

\vspace{1em}
\begin{tcolorbox}[question]
\textit{What advice would you give on selecting model design choices such as architecture, tokenization, modality and scale?}
\end{tcolorbox}
\vspace{1em}

Governments can follow existing, proven technical paths that have been demonstrated in published papers, technical reports and successful open models, rather than pursuing uncertain architectural innovations. This can significantly reduces unnecessary technical risk, cost and trial-and-error overhead.

\vspace{1em}
\begin{tcolorbox}[question]
\textit{Were there unexpected difficulties in obtaining or collecting data?}
\end{tcolorbox}
\vspace{1em}

\subsubsection*{Singapore: SEA-LION}
Yes, data acquisition proved more complex than anticipated. First, many domains that require culturally-specific context and locally grounded data, such as agriculture in rural areas, lived experiences of migrant workers and other areas where social norms need to be observed, were unavailable online and had to be collected directly through field interviews. Second, getting high-quality data exhibits persistent challenges. 
Synthetic data offered efficiency, but introduced trust and validation issues, requiring manual verification. Third, copyright constraints raised costs significantly, especially following high-profile lawsuits that increased data owners’ price expectations.

\subsubsection*{Switzerland: Apertus}

We compiled our own multilingual training and evaluation datasets, relying exclusively on public, versioned data sourced from web crawls. While this choice simplified data access, data curation emerged as a major challenge. Ensuring compliance with our data policies required extensive filtering and validation, and the process of defining and refining those policies itself was time-consuming.

\vspace{1em}
\begin{tcolorbox}[question]
\textit{Looking back, what lessons would you highlight for governments considering similar building efforts?}
\end{tcolorbox}
\vspace{1em}

Regarding model building, full pretraining is extremely costly and talent-intensive. When resource or expertise are limited, focusing on post-training or alignment on existing open base models is often a more practical choice.
Regarding data building, governments should avoid collecting data from scratch blindly. Carefully assess what can be reliably generated through synthetic methods to avoid unnecessary costs.

\vspace{1em}
\begin{tcolorbox}[question]
\textit{What did ``good performance'' mean in your context?}
\end{tcolorbox}
\vspace{1em}

\subsubsection*{Switzerland: Apertus}

The notion of ``good performance'' for a government sovereign LLM is not universal. It depends critically on the specific objectives, constraints and societal values of the deploying institution. 

In our context, ``good performance'' meant demonstrating that responsible AI development does not entail a significant capability trade-off. 
Concretely, Apertus aimed to remain within striking distance of state-of-the-art open models on standard benchmarks, while excelling on multilingual evaluations that reflect cultures and languages traditionally underrepresented in LLM development - and it managed to achieve leading performance across many individual languages. 
More importantly, its distinguishing contribution lies in its commitment to responsibility: full openness, transparent design, respect for data ownership, and broad linguistic coverage. 
Our goal was not to dominate existing leaderboards, but to establish a credible, high-performing foundation for an alternative paradigm of AI development that prioritizes responsibility, inclusivity, and global representation.

\vspace{1em}
\begin{tcolorbox}[question]
\textit{Were there any outcomes that surprised you?}
\end{tcolorbox}
\vspace{1em}

\subsubsection*{Singapore: SEA-LION}
Yes. By involving partner countries such as Thailand and Indonesia directly in data building and model co-training, SEA-LION evolved into a model for inclusive and community-driven AI development. This collaborative development enabled more localized variants of SEA-LION in Southeast Asia region through continued model adaptation, which allows SEA-LION to grow organically and function as a shared regional AI infrastructure. Beyond Southeast Asia, this model had also attracted interest globally from other parts of Asia, Africa and some NGOs, who see it as a promising example of culturally grounded and participatory AI development.

%% file: sections/conclusions.tex
\section{Conclusions}

The choice between buying or building a language model defines how a nation delivers the upsides of AI technologies to its citizens. While buying a model provides a quick start with less technical effort, it often creates a long term dependence on external companies and their rules. Building a sovereign model is more difficult and requires more resources, but it allows a country to own its technology and stay true to its own values and laws.

As this framework shows, these LLMs are more than just software; they are a new kind of public utility. There is no single right answer for every country. Instead, each government must weigh the trade-offs based on its own budget, talent, and vision for the future. By looking at all these factors, leaders can choose the best way to provide AI as a secure, useful, and sustainable resource for their people.

\section*{Acknowledgments}
We sincerely thank Jian Gang Ngui and Leslie Teo from AI Singapore for their thorough review of this paper and for providing insightful and constructive suggestions that significantly improved its clarity and quality.
{We also thank the ACM reviewers and ACM Technology Policy Council members for their careful review and valuable feedback, which helped further improve the quality of this work.}